\documentstyle[preprint,prl,floats,aps]{revtex}

\begin{document}
\draft

\title{The Structural Glass Transition \\
and the Entropy of the Metastable States}

\author{R\'emi Monasson $^{1,2}$}
\address{$^1$ INFN and Dipartimento di Fisica, P.le Aldo Moro 2,
I-00185 Roma, Italy\\
$^2$ Laboratoire de Physique Th\'eorique de l'ENS \footnote{Unit\'e propre
du CNRS, associ\'ee \`a l'ecole Normale Sup\'erieure et \`a
l'Universit\'e de Paris Sud}, 24 rue Lhomond, F-75231
Paris cedex 05, France}

\maketitle
\begin{abstract}
The metastable states of a glass are counted by adding a weak pinning
field which explicitly breaks the ergodicity.  Their entropy, that is
the logarithm of their number, is extensive in a range of temperatures
$T_G < T < T_C$ only, where $T_G$ and $T_C$ correspond to the ideal
calorimetric and kinetic glass transition temperatures respectively.
An explicit self-consistent computation of the metastable states
entropy for a non disordered model is given.
\end{abstract}


\narrowtext

During the last decade, a great deal of work has been devoted to the
understanding of the glass transition. Basically, two different
approaches have been employed to tackle this problem. On the one hand,
the density functional theory (DFT) \cite{dft} is an equilibrium
approach where the static density field obtained through the
minimization of a free-energy functional becomes inhomogeneous below
the structural glass transition temperature $T_G$.  On the other hand,
the mode coupling theory (MCT) \cite{mct} is an off-equilibrium theory
which self-consistently treats the microscopic dynamical correlation
and response functions in the liquid phase. The onset of the glassy
state is then due to a kinetic transition at a temperature $T_C$
higher than $T_G$. The breaking of ergodicity at $T_C$ coincides with
the appearance of non vanishing density fluctuations in the long time
limit, the system becoming partially frozen in metastables states with
very large relaxation times.

The connection between the dynamic and the static approaches can been
made more transparent and rigorous in the special case of mean-field
spin glasses \cite{kirwol,kirthi}. It has indeed been found that the
models exhibiting a discontinuous replica symmetry breaking \cite{mpv}
at the static transition temperature $T_S$ present a dynamic
transition at a higher temperature $T_D$ below which equilibrium is
never reached \cite{kirthi,kirwol,margin}. In addition, the
temperature $T_D$ whose significance is a priori purely dynamical may
be computed from the Gibbs partition function using the so-called
``marginality condition'' \cite{margin}.  This is related to the
intuitive feeling that the dynamical transition should arise when the
free--energy landscape becomes ``rough'' and includes many states
separated by extensive barriers while at the same time the replica
symmetry remains unbroken ($T_D >T_S$).  Recently, it has been shown
that these results may be extended to mean--field spin models with
complicated but non--random interactions \cite{boumez}. A glassy
system, once trapped in a metastable state, is indeed partially frozen
and its slow degrees of freedom may act as self-induced ``quenched''
constraints with respect to the fast ones \cite{mct,kirthi,boumez}.

In this letter, we wish to investigate the metastable states in the
equilibrium free-energy landscape in any glassy system with or without
quenched disorder. More precisely, we shall define the entropy ${\cal
S}_{hs}$ as the logarithm of the number of these ``hidden''
metastable states and present a general scheme to compute this
quantity.  It will be shown that for mean-field disordered models
${\cal S}_{hs}$ is extensive in the range of temperatures $T_S < T <
T_D$ only, and gives also some important information about the
structure of the states below $T_S$. We shall argue that the same
holds in the absence of disorder and that these two temperatures then
correspond to the ideal calorimetric and kinetic temperatures $T_G$
and $T_C$ respectively.  The replica formalism will then be adapted to
the case of systems without disorder to derive their metastable states
entropy ${\cal S}_{hs}$ and the glassy correlation functions inside
these states.  An example of such a calculation will be given for a
simple model.

Let us start with a theory of a field $\phi (x)$ defined by a
Hamiltonian $H [\phi]$.  For simplicity, we shall use scalar notation
though $x$ spans a $D$-dimensional space and $\phi (x)$ can be an
$N$-component field. The equilibrium Gibbs free--energy at the
temperature $T=\frac{1}{\beta}$ is given by
\begin{equation}
F_{\phi} (\beta)=-\frac{1}{\beta} \log \int d\phi (x) \ e^{ -\beta
H [\phi ]}
\label{fphi}
\end{equation}
For a usual ferromagnet, the emergence of a
spontaneous magnetization at low temperature coincides with the
breaking of the symmetry $\phi(x) \to - \phi(x)$ of the
Hamiltonian. Below the Curie temperature, the physical decomposition
of the Gibbs free--energy into two states of opposite magnetizations
may be obtained by imposing a small (but finite) external field
aligned along the up or down directions. In the case of disordered
systems or glasses, there is however no a priori privileged direction
towards which the field $\phi (x)$ will point once stuck in a
metastable state.  We can nevertheless choose a possible direction,
given by another field $\sigma (x)$, and compute the free--energy of
our system when it is weakly pinned by this external quenched field
\begin{equation}
F_{\phi}\left[\sigma ,g,\beta\right] = -\frac{1}{\beta} \log
\int d\phi (x) \ e^{ -\beta H [\phi ] - \frac{g}{2}
\int\!\! dx ( \sigma (x) - \phi (x))^2}
\label{fphig}
\end{equation}
where $g>0$ denotes the strength of the coupling. This free-energy
(\ref{fphig}) will be small when the external perturbing field $\sigma
(x)$ lies in a direction corresponding to the bottom of a well of the
unperturbed free-energy (\ref{fphi}).  Therefore, we should be able to
obtain useful information about the free-energy landscape by scanning
the entire space of the configurations $\sigma (x)$ to locate all
the states in which the system can freeze after spontaneous
ergodicity breaking ($g\to 0$). According to this intuitive idea, we now
consider the field $\sigma(x)$ as a thermalized variable with the
``Hamiltonian'' $F_{\phi} \left[\sigma ,g,\beta\right]$.
The free-energy of the field $\sigma$ at inverse temperature
$\beta m$ where $m$ is a positive free parameter therefore reads
\begin{equation}
F_{\sigma}(m,\beta) = \mathop{\rm lim}_{g \to 0^+} -\frac{1}{\beta m}
\log \int d\sigma (x)\ e^{-\beta m F_{\phi}[\sigma ,g,\beta ]}
\label{fsigm}
\end{equation}
When the ratio $m$ between the two temperatures
is an integer, one can easily integrate $\sigma (x)$ in
eqn.(\ref{fsigm}) after having introduced $m$ copies $\phi ^{\rho}
(x)$ ($\rho =1...m$) of the original field to obtain the relation
\begin{equation}
F_{\sigma}(m,\beta) = \mathop{\rm lim}_{g \to 0^+}
-\frac{1}{\beta m}\log \int \prod _{\rho =1}^m
d\phi^{\rho} (x) \ e^{ -\beta \sum _{\rho } H [\phi ^{\rho}] +
\frac{1}{2} \sum _{\rho,\lambda} g^{\rho \lambda}\int dx \phi ^{\rho}
(x) \phi ^{\lambda} (x)}
\label{frepl}
\end{equation}
where $g^{\rho \lambda}=g(\frac{1}{m}-\delta^{\rho \lambda})$.  Let us
define two more quantities related to the field $\sigma$~: its
internal energy $W(m,\beta )=\frac{\partial (m F_{\sigma})} {\partial
m}$ and its entropy $S(m,\beta)=\beta m^2 \frac{\partial
F_{\sigma}}{\partial m}$.  Since the case $m=1$ will be of particular
interest, we shall use hereafter $F_{hs} (\beta) \equiv
W(m=1,\beta)$ and ${\cal S}_{hs}  (\beta) \equiv S(m=1,\beta)$ where
$hs$ stands for ``hidden states''.  We stress that $S(m,\beta)$ and
$\beta ^2 \frac{\partial F_{\phi} }{\partial \beta}$ which are
respectively the entropies of the fields $\sigma$ and $\phi$ are two
distinct quantities with different physical meanings.

When the pinning field $\sigma (x)$ is thermalized at the same
temperature as $\phi (x)$, that is when $m=1$, one sees from
eqn.(\ref{frepl}) that $F_{\phi} (\beta)=F_{\sigma}(m=1,\beta)$.
The basic idea of this letter is to decompose $F_{\sigma}$
into its energetic and entropic contributions to obtain
\begin{equation}
{\cal S}_{hs}(\beta)=\beta \bigg[ F_{hs}(\beta) - F_{\phi}
(\beta) \bigg]
\label{fond}
\end{equation}
To get some insights on the significance of the above relation, we
shall now turn to the particular case of disordered mean-field
systems.  We shall see how it rigorously gives back some analytical
results derived within the mean-field TAP and dynamical approaches
\cite{kirwol,margin,marc,giorgio,tap}. We shall then discuss the physical
meaning of identity (\ref{fond}) for the general case of glassy systems.

The free-energy $F_{\phi} (\beta)$ of a mean-field disordered system
is a self-averaging quantity which may be computed using the replica
trick \cite{mpv} to end up with $F_{\phi}(\beta)= {\displaystyle
\mathop{\rm lim}_{n \to 0}\mathop{\rm Ext}_{\{q ^{ab}
\}}{\cal F}_{\phi} (\{q ^{ab} \})}$. $q^{ab}=\frac{1}{V}\int dx
\phi ^a(x)\phi ^b(x)$ are the overlaps between the $n$ replicas
($V\equiv \int dx$ is the volume of the system). Above the static
transition temperature $T_S$, the physical saddle-point of ${\cal
F}_{\phi}$ is replica symmetric (RS) $q^{a\ne b}=q$ (for simplicity a
spherical normalization $q^{aa}=1$ has been assumed).  If we now
compute the free--energy (\ref{fsigm}) of the field $\sigma$ by
introducing $n$ replicas and averaging over the quenched disorder, we
obtain from eqn.(\ref{frepl}) the same free-energy functional ${\cal
F}_{\phi}$ where the number of replicas $\phi ^{a\rho}(x)$ now equals
$n\times m$ and with the additional term $\frac{V}{2}\sum _{a=1}^n
\sum _{\rho,\lambda =1} ^m g^{\rho \lambda} q^{a\rho , a\lambda}$.
This interaction term explicitly breaks the symmetry of permutations
of the $n\times m$ replicas \cite{parvir} into $n$ groups of $m$
indistinguishable replicas. Consequently, even above $T_s$, the
simplest Ansatz one can resort to contains at least one step of
replica symmetry breaking (RSB) which reads
\begin{equation}
F_{\sigma} (m,\beta) =
\mathop{\rm Ext}_{q_0, q_1}\ {\cal F}_{\phi} (q_0,q_1,m,\beta)
\label{desor}
\end{equation}
when $g\to 0^+$.  As a result of the introduction of the field
$\sigma$, we have obtained the usual one--step expression of the
free--energy but without any optimization over the free parameter $m$
which we can choose at our convenience (see \cite{reti} for a related
case where $\sigma$ acquires a simple geometrical interpretation).
Let us now send $m\to 1$ while $T>T_S$. The saddle-point equation over
$q_0$ becomes identical to the RS equation for $q$. Thus $q_0=q$ and
we find $F_{\sigma}(m=1,\beta)=F_{\phi}(\beta)= \mathop{\rm Extr}_{q}
{\cal F}_{\phi} (q,\beta)$ as expected. Defining ${\cal V}(q_1) \equiv
\beta {\displaystyle \frac{\partial {\cal F}_{\phi} }{\partial m}
(q_0=q,q_1,m=1,\beta)}$, the optimization condition over the second
overlap implies that $q_1$ must be a stable local minimum of ${\cal
V}$. The entropy of $\sigma$ is then ${\cal S}_{hs}={\cal
V}(q_1)$. It turns out that ${\cal V}$ defined above is equal to the
potential recently introduced in mean-field glasses
\cite{giorgio,silvio} to compute the marginality temperature $T_D$ at
which the relaxational dynamics exhibits a drastic slow down
\cite{kirwol,margin} as mentioned in the introduction.  The typical
behaviour of the entropy ${\cal S}_{hs}(\beta)$ is as follows.  At high
temperature, there is only the RS solution $q_1=q, {\cal S}_{hs}=0$.
At a given $T_D$, there appears a non trivial saddle-point $q_1$
(which shifts the free--energy (\ref{desor}) by an extensive amount of
order $g V$ \cite{parvir}) and the entropy ${\cal S}_{hs}$ shows a
first order jump. When $T$ decreases, ${\cal S}_{hs}$ goes down and
vanishes at $T_S$. Note that ${\cal S}_{hs} (T=T_S)=0$ is
mathematically equivalent to the usual optimization condition of the
one-step free--energy with respect to $m$. This is an important remark
we shall comment on in the following.

The marginality condition \cite{margin,giorgio} tells us that $T=T_D$
is the temperature of the onset of an exponential number of such
metastables states. Using the TAP formalism \cite{tap}, one can indeed
write the set of equations that the mean-field local magnetizations
$\{M(x)\}$ must satisfy. The paramagnetic solution $M(x)=0$ (for
simplicity we suppose that $q=(\frac{1}{V} \int dx M(x))^2$ vanishes
in the RS phase) gives the RS free--energy density as expected. Below
$T_D$ the RS free--energy density is also achieved by the sum of an
exponential number of solutions with a higher free--energy and an
internal overlap $q_1=\frac{1}{V}\int dx M(x)^2$ \cite{kirwol}.  It
has been recently verified in the particular case of the random
orthogonal model \cite{marc} that the logarithm of the number of these
solutions \cite{bm} - which is sometimes called complexity or
configurational entropy \cite{kirwol,kauz} - coincides with the
entropy ${\cal S}_{hs}$ defined in this letter.

The significance of formula (\ref{fond}) may now be discussed for
glasses in general. At high temperature, the pinning due to $\sigma$
is not sufficient to break ergodicity~:
$F_{hs}(\beta)=F_{\phi}(\beta)$ and ${\cal S}_{hs}(\beta) =0$.
Physically, there exists only one state. When the temperature goes
down, there may appear some high barriers which separate an
exponential number of metastable states of free--energies higher than
the true $F_{\phi}(\beta)$. As long as the number of these states does
not compensate for their small weights, they are not ``seen'' by the
Gibbs partition function. At some temperature $T_D$, their number
$e^{{\cal S}_{hs}}$ becomes large enough to make up for the difference
of free--energies and the identity (\ref{fond}) expresses this
compensation mechanism. In a system with finite range interactions,
the ergodicity breaking taking place at $T_D$ is not complete. The
partial freezing of collective modes inside the metastable states
makes the dynamics very slow but does not forbid some microscopic
changes (the so-called activated processes) \cite{kirwol}. Thus, no
discontinuties occur at $T_D$ for the thermodynamic quantities and it
is reasonable to think that $T_D$ coincides with the ideal kinetic
transition temperature $T_C$ for glassy systems.  Below $T_D$,
$F_{\phi}(\beta )$ comes from the superposition of many states with
high free--energies. The number $e^{{\cal S}_{hs}}$ of these hidden
states decreases since their free--energy $F_{hs}$ gets closer and
closer to the true value $F_{\phi}$. At a given temperature $T_S$,
these states cease to be metastable since their free--energy $F_{hs}$
equals $F_{\phi}$, implying from (\ref{fond}) that the entropy ${\cal
S}_{hs}$ is not extensive anymore \cite{kauz}. The true
thermodynamical transition therefore takes place at $T_S$ which
corresponds to the usual calorimetric glass transition temperature
$T_G$. Below $T_S$, formula (\ref{fond}) cannot hold any longer since
it would predict a negative configurational entropy of metastable
states having a lower free--energy than $F_{\phi}$.  Physically, one
expects that freezing into a small (non exponential) number of states
will still occur and that ${\cal S}_{hs} (T<T_S)=0$. As a consequence,
the effective temperature of the states, that is of the field
$\sigma$, becomes higher than the temperature of the field $\phi$ and
$m$ is determined through the condition that the hidden states'
entropy vanishes. If one goes back to the case of mean-field
disordered systems, one finds that $m$ is such that the one-step
free--energy derivative with respect to $m$ equals zero.  Therefore,
we have obtained a simple physical interpretation of the usual
optimization condition of ${\cal F}_{\phi}$ with respect to $m$ in the
RSB phase, whose meaning has always been far less clear than the
optimization with respect to the overlaps $q_0$ and $q_1$ \cite{mpv}.
Furthermore, let us notice that if ${\cal S}_{hs}$ is already non
extensive at $T_D$ then $T_D$ and $T_S$ must coincide. This is what
happens for systems with a continuous RSB transition \cite{mpv} where
the intermediate phase $T_S < T < T_D$ is skipped. Such systems seem
therefore to exhibit a less generic behaviour \cite{kirwol,kirthi}.

Formula (\ref{frepl}) is a convenient starting point to compute ${\cal S}
_{hs}$ in systems without quenched disorder.
We begin with $m$ uncoupled copies $\phi ^{\rho} (x)$
of the model. The matrix of the correlation functions of the
global system is then a priori diagonal~: $G^{\rho \lambda}(x-y) \equiv
\langle \phi ^{\rho} (x)\phi ^{\lambda} (y)\rangle = \delta ^{\rho \lambda}
G(x-y)$. For simplicity we shall assume that in the liquid phase the
average value of the field $\langle \phi(x)\rangle$ is equal to zero.
The onset of the glassy phase will be characterized by the appearance
of metastable states in which the expectation value
$\langle \phi(x)\rangle$ does not vanish anymore and therefore by the
emergence of non zero off-diagonal propagators
\begin{equation}
G_{hs} (x-y) = \langle \phi(x)\rangle \langle\phi(y)\rangle =
\langle \phi ^{\rho} (x)\phi ^{\lambda} (y)\rangle \quad (\rho \ne \lambda)
\label{glass}
\end{equation}
in the limit $m\to 1$ \cite{bre}. As discussed above, the diagonal correlation
function $G(x-y)$ is simply obtained from the calculation of $F_{\phi}
(\beta)$ for a single system ($m=1$). The computation of $F_{\sigma}(m
\ne 1,\beta)$ is more delicate since a perturbative expansion of the original
Hamiltonian will never be able to generate an effective coupling
between the replicas and to detect the first order glass transition.
We shall now present a simple self-consistent calculation which permits us
to obtain ${\cal S}_{hs}$ and $G_{hs} (x-y)$.

Let us consider a system of $N$ particles at positions $x_i$
($i=1...N$) interacting through a two--body potential $\sum _{i<j}
V(|x_i - x_j|)$.  In the grand-canonical ensemble, this system is
described by the Hamiltonian $H[\phi ] = \frac{1}{2} \int dx dy\ \phi
(x) V^{-1} (x-y) \phi (y) - \mu \int dx \exp{i \phi(x)}$ where $\mu$
is the chemical potential and $\phi (x)$ the conjugated field to the
density of particles at point $x$. To obtain an analytically tractable
problem, we shall make two simplifications. First, while the previous
general expression contains interactions at all orders in $\phi$, we
shall keep only the quartic vertex. Secondly, we shall consider a
$O(N)$ version of the above Hamiltonian and compute the free--energy
$F_{\sigma}(m,\beta )$ using Bray's self-consistent screening
approximation \cite{bray}. The expansion parameter $1/N$ plays the
role of the chemical potential $\mu$.  We do not expect this model to
exhibit a glass transition for large $N$ but the small $N$ case may
have an important physical interest, e.g. polymers which are related
to $N\to 0$ \cite{deg} are known to have a glassy behaviour at low
temperatures.  Bray's calculation is exact to order $1/N$ and contains
a partial resummation of an infinite class of diagrams \cite{bray} and
is thus well defined in the whole range $0<N < \infty$. With the
Ansatz $G^{\rho \lambda}(k)=\delta ^{\rho \lambda} G(k) + (1-\delta
^{\rho \lambda}) G_{hs}(k)$ \cite{bre}, one obtains the free--energy
$F_{\sigma}(m,N)$ as an extremum over the set of all propagators
$G(k)$ and $G_{hs}(k)$ \cite{mezyou}. When $m\to 1$, the diagonal
propagators are solutions of the implicit equations
\begin{equation}
\frac{1}{G(k)}=\frac{1}{V(k)}+\int dq\ G(q) + \frac{2}{N} \int dq\ \frac
{G(k-q)}{1+\Pi (q)}
\label{eqbr}
\end{equation}
where $\Pi(k)\equiv \int dq G(k-q)G(q)$. Eqn.(\ref{eqbr}) is
identical to Bray's result as expected \cite{bray}. The entropy of the
metastable states then reads
\begin{equation}
{\cal S}_{hs}(N) = \mathop{\rm Ext} _{G_{hs}(k)}  \int\!\!dk \left[
s \left(\frac{G_{hs}(k)}{G(k)}\right) - \frac{1}{N}
s \left(\frac{\Pi _{hs}(k)}{1+\Pi (k)}\right) \right]
\label{phi4}
\end{equation}
where $\Pi _{hs}(k)\equiv \int dq G_{hs}(k-q)G_{hs}(q)$ and
$s(x)\equiv - x -\log(1-x)$. The numerical resolution of the
saddle-point equations steming from eqn.(\ref{eqbr},\ref{phi4}) in
dimension $D=3$ shows that a non zero set of propagators $G_{hs}$
may appear when $N$ becomes lower than a given $N_C$ which depends on
the bare propagator $V(k)$. Despite different choices of $V(k)$, we
have always found $N_C<1$. A more careful analysis of the equations
would however be useful.  In dimension $D=0$, the equations for $G$
and $G_{hs}$ may be solved exactly and we find that there exists a
first order transition at some small enough critical value of $N$. One
can show that $N_C$ is always lower than one. If for instance we
choose the bare mass $V=-0.2 $, ${\cal S}_{hs}$ is equal to zero
when $N > N_C \simeq 0.65$ and jumps discontinuously to 0.31 at the
transition with $G_{hs}/G \simeq 0.91$.  The entropy then decreases
smoothly when $N$ gets smaller and vanishes at $N _G \simeq 0.54$
where $G_{hs}/G \simeq 0.96$. We notice that the ratios $N_C /N _G$
and $G_{hs}/G$ are remarkably similar to the values of $T_D/T_S$ and
$q_1$ which may be found in mean-field disordered models
\cite{margin,boumez,marc}. Though one must consider this result with
caution due to the approximations made in its derivation, it seems
that the free--energy landscape of the $(\vec \phi ^2)^2$ model may be
complicated at small $N$ ($N<1$), even above the ferromagnetic
transition temperature.  This is strongly reminiscent of the random
field Ising model which is recovered here if $N=1$ and $g$ keeps a
finite value \cite{mezyou}.

A deeper and more rigorous understanding of the mechanism of the
ergodicity breaking occurring at $T_C$ is still to be found. In this
respect, one could try to transpose the TAP approach to non disordered
models. Let us call $\Gamma (M)= g\int dx \sigma (x) M(x) +
F_{\phi}[\sigma,g,\beta]$ the Legendre transform of the
``Hamiltonian'' of the field $\sigma$. The free--energy $F_{\phi}
(\beta)$ is equal to the minimum of the effective potential $\Gamma$
at $M(x)=\langle\phi(x)\rangle$. It may well happen that at
$T_C$ there appear many different minima of $\Gamma$ while at higher
temperatures the only solution is $M(x)=0$.  If so, the presence of
the source $\sigma$ would lift this degeneracy by selecting the
closest solution $M(x)\ne 0$ (i.e. having the largest overlap with
$\sigma $). The propagator $G_{hs}(x-y)$ would then be the average
of $M(x)M(y)$ over all these solutions, that is over all possible
fields $\sigma$. A natural idea is thus to define a partition function
$Z_{solutions}$ as the sum of the Gibbs weights $e^{-\beta \Gamma
(M)}$ of all the solutions of $\frac{\partial \Gamma }{\partial M(x)
}=0$.  It would be very interesting to confirm or disprove that the
equality between $Z_{solutions}$ and the partition function of the
field $\phi$ eqn.(\ref{fphi}) also holds in the glassy phase below
$T_C$.

We would like to thank M. Ferrero, S. Franz, D.J. Lancaster,
G. Parisi, M. Potters, M. Virasoro and R. Zecchina for numerous and
fruitful discussions. We are grateful to S. Franz and G. Parisi for the
communication of some of their results \cite{silvio} before publication.

\end{document}